\documentclass[aps,prc,preprint,groupedaddress]{revtex4-1}

\usepackage{graphicx}
\usepackage{amsmath}
\usepackage{bm}

\begin{document}


\title{{\it Ab initio} perturbation calculations of realistic effective interactions in the Hartree--Fock basis}


\author{Q. Wu}
\author{F. R. Xu}
\email{frxu@pku.edu.cn}
\author{B. S. Hu}
\author{J. G. Li}
\affiliation{School of Physics, and State Key Laboratory of Nuclear Physics and Technology, Peking University, Beijing 100871, China}


\date{\today}

\begin{abstract}
We perform two types of {\it ab initio} perturbation calculations of effective interactions in the Hartree--Fock (HF) basis instead of the harmonic-oscillator basis: one is called the Brillouin--Wigner (BW) perturbation and another is called the Rayleigh--Schr\"odinger (RS) perturbation. It is shown that the HF basis provides good convergences. We also benchmark the perturbation calculations with the in-medium similarity renormalization group (IM-SRG) which is a nonperturbative method. In the HF basis some-type perturbation diagrams can be cancelled out, while the cancellation does not happen in the harmonic-oscillator basis. We have investigated the {\it sd} shell using the chiral N$^3$LO potential softened by $V_{\text{low-}k}$. With the low-momentum N$^3$LO potential, we first perform the spherical HF calculations for the $^{16}$O core of the {\it sd} shell, and the realistic effective two-body interactions for the HF {\it sd}-shell space can be derived by the perturbation calculations. The calculations give simultaneously effective single-particle energies and excitation spectra of two valence particle systems (i.e., $^{18}$O, $^{18}$F, and $^{18}$Ne in the {\it sd} shell). Convergences have been analyzed order by order. The perturbation calculations are in fairly good agreement with nonperturbative method IM-SRG. We find that the HF RS perturbation gives even better results compared with the BW perturbation. The HF realistic effective interactions derived by the perturbations can be used for further shell-model calculations.

\end{abstract}


\maketitle

\section{Introduction}
Based on nuclear force from chiral effective field theory \cite{RevModPhys.81.1773,MACHLEIDT20111}, and low-momentum interactions under the renormalization philosophy \cite{BOGNER201094}, tremendous progress of {\it ab initio} nuclear many-body calculations have been achieved in recent years. Different many-body methods, such as self-consistent Green's function method \cite{PhysRevC.89.061301,PhysRevLett.111.062501,PhysRevC.87.011303}, coupled cluster (CC) theory \cite{PhysRevC.69.054320,RevModPhys.79.291,PhysRevC.82.034330,PhysRevLett.92.132501,BINDER2014119} and in-medium similarity renormalization group (IM-SRG) \cite{PhysRevC.87.034307,PhysRevLett.106.222502,HERGERT2016165}, have been successfully applied to study both closed- and open-shell nuclei in medium-mass region, yielding similar results with same input nuclear Hamiltonians \cite{HERGERT2016165,PhysRevLett.117.052501}. However, the direct calculations are limited to nuclei with few valence nucleons. For open-shell system with many valence nucleons, the shell model remains powerful. One feasible way for these open-shell {\it ab initio} calculations is to follow the shell-model paradigm by constructing the effective interaction from first principles. With this same goal, valence nucleon effective shell-model interactions have been introduced based on the CC method \cite{PhysRevLett.113.142502}, IM-SRG method \cite{PhysRevLett.113.142501,Tsukiyama2012}, as well as no-core shell model (NCSM) \cite{PhysRevC.91.064301,PhysRevC.78.044302,0954-3899-36-8-083101}.

While all above calculations are nonperturbative, there has been a lot of earlier work calculating the effective shell-model interaction perturbatively within the framework of many-body perturbation theory (MBPT) \cite{BRANDOW1967,Kuo1974,RevModPhys.49.777}. MBPT has been successful in deriving model-space effective interactions from realistic nuclear forces \cite{HJORTHJENSEN1995125,1528474b45a64fbf9653713ca75968c0,Coraggio2009b,Coraggio2012}. The perturbation approach separates a core and all computed observables are relative to the core \cite{BRANDOW1967,RevModPhys.49.777}. In most cases, harmonic-oscillator (HO) Hamiltonian has been chosen as the unperturbed Hamiltonian and many-body configurations are constructed in the HO basis. However, the computations within the HO basis can be dependent on the choice of the basis parameter value, $\hbar\omega$ \cite{Coraggio2012,Tsukiyama2012,HJORTHJENSEN1995125}.

In contrast to the HO basis, the self-consistent HF basis provides improved convergences of perturbation calculations \cite{TICHAI2016283,Hu2016}. In addition, in the HF basis some-type perturbation diagrams can be cancelled out, while the cancellation does not happen in the harmonic-oscillator basis. There have already been calculations of realistic effective interaction within the HF basis using MBPT. In Ref. \cite{CORAGGIO200543}, within the self-consistent HF basis, the $p$-shell effective interaction and single-particle energies (SPEs) were obtained using the perturbation method with the $V_{\text{low-}k}$ CD-Bonn $NN$ potential, giving satisfactory results for Li, Be, B isotopes compared with the experiment data. In Ref. \cite{Tsukiyama2012}, the $sd$-shell nucleus $^{18}$O has been studied, using IM-SRG and MBPT with the HO and HF basis, showing that the HF MBPT gives similar results to the IM-SRG calculations. The HF MBPT calculations are almost independent on the parameter $\hbar\omega$, while the HO MBPT results are sensitive to the $\hbar\omega$ value \cite{Tsukiyama2012}. Therefore, the perturbation approximation in the HF basis provides a promising method to derive the realistic effective nucleon-nucleon interaction and SPEs for many-body calculations.

In this paper, we present a detailed investigation  of the HF-basis perturbation calculations for the $sd$ shell. The order-by-order convergence property is analysed. We compare the obtained SPEs, nuclear excitation spectra, and effective interaction matrix elements with those calculated by the nonperturbative method IM-SRG with the same input Hamiltonian. We perform two different perturbation calculations by so-called Brillouin--Wigner (BW) and Rayleigh--Schr\"odinger (RS) perturbations. In the BW calculations, the perturbative expansion of the $\hat{Q}$-box is calculated up to third order, and the energy-dependent secular equation is solved by iterations. While in the RS calculations, the perturbative expansion is calculated up to second order.

\section{Theoretical Frameworks}
\subsection{Perturbation Frameworks}
The basic idea of deriving the effective interaction is to decouple the interested model-space Hamiltonian from the rest of the full Hamiltonian by similarity transformations. The derived effective Hamiltonian defined in the model space should reproduce the low-energy eigenvalues of interest of the original Hamiltonian. The decoupling can be carried out by perturbation expansions \cite{Kuo1974,Hjorth-Jensen1995,Coraggio2012}, or introducing configuration truncations in nonperturbative methods \cite{Tsukiyama2012,PhysRevLett.113.142501}.

The MBPT formulation has been in detail described in, e.g., \cite{Hjorth-Jensen1995,Dean2004,Coraggio2009b,Coraggio2012,Takayanagi2011a}. Here we briefly outline the approach starting from the Bloch-Horowitz (BH) effective Hamiltonian in the framework of the time-independent perturbation theory \cite{BRANDOW1967,RevModPhys.49.777}.

The nuclear system is described by the Schr\"odinger equation
\begin{eqnarray}
H |\Psi_\lambda\rangle=E_\lambda |\Psi_\lambda\rangle,
\end{eqnarray}
where $H$ is the Hamiltonian, and $|\Psi_\lambda\rangle$ is the eigenvector corresponding to the eigenvalue $E_\lambda$. The BH effective Hamiltonian is defined as
\begin{eqnarray}
H_{\text{BH}}(E) = PHP + PHQ\frac{1}{E-QHQ}QHP,
\label{eq:BH}
\end{eqnarray}
where $P$ and $Q$ are the projection operators onto the model space and its complement. The BH effective Hamiltonian is energy-dependent, and it gives the same solutions as the original Hamiltonian $H$ through the energy-dependent secular equation,
\begin{eqnarray}
H_{\text{BH}}(E_\lambda) |\phi_\lambda\rangle=E_\lambda |\phi_\lambda\rangle,
\label{eq:BHeq}
\end{eqnarray}
where $|\phi_\lambda\rangle=P|\Psi_\lambda\rangle$  is the  P-space projection of the eigenvector corresponding to the eigen energy $E_\lambda$.

In the perturbation expansion, the full Hamiltonian $H$ is split into two parts,
\begin{eqnarray}
H&=&H_0+H_1,
\end{eqnarray}
where $H_0$ is the unperturbed Hamiltonian and $H_1$ is the residual interaction treated as the perturbation. The model space is usually chosen as the space spanned by a set of eigenstates of $H_0$. When applied to an $A$-nucleon system, the unperturbed Hamiltonian is chosen, in the second-quantized formulation, as
\begin{eqnarray}
H_0=\sum\limits_i \epsilon_i\hat{a}_i^\dagger \hat{a}_i,
\end{eqnarray}
where $\epsilon_i$ and $\hat{a}_i^\dagger$ are the energy and the creation operator of the single-particle (s.p.) state labelled by $i$, respectively. With the s.p. basis states, we can construct many $A$-body Slater determinants as the configurations. The configuration space is divided into model space and its complement. A reference Slater determinant of a close shell core with $A_\text{c}$ nucleons is chosen, whose orbits are always occupied by the configurations in the model space. The orbits occupied (unoccupied) by the core are called hole (particle) states. The s.p. states can be divided into two categories: valence states and passive states. Passive states are the s.p. states that are always occupied or unoccupied in the model space, while the rest $n=A-A_\text{c}$ valence nucleons are distributed among the valence states.

An important step in the derivation of the effective interaction is the factorization of the core as a result of the factorization theorem \cite{BRANDOW1967,Kuo1974,RevModPhys.49.777}, which enables us to directly calculate the relative energy $E^{(\nu)}=E-E_\text{c}$ ($E_\text{c}$ is the binding energy of the core nucleus), and the $(A_\text{c}+n)$-body problem is reduced to a $n$-body problem \cite{BRANDOW1967,Kuo1974,RevModPhys.49.777}. After the procedure, Eqs.~(\ref{eq:BH})-(\ref{eq:BHeq}) are reduced to a form with only valence nucleons involved \cite{BRANDOW1967,Kuo1974,RevModPhys.49.777},
\begin{eqnarray}
  &H_{\text{eff}}(E)=PH_0P + \hat{Q}(E),
  \label{eq:HeffE}
\end{eqnarray}
and
\begin{eqnarray}
&H_{\text{eff}} (E^{(\nu)}_\lambda) |\chi_\lambda\rangle=E^{(\nu)}_\lambda |\chi_\lambda\rangle,
\label{eq:secular}
\end{eqnarray}
where the so-called $\hat{Q}$-box is defined as $\hat{Q}(E)=PH_1P+PH_1Q\dfrac{1}{E-QHQ}QH_1P$ in Eq.~(\ref{eq:HeffE}). $PH_0P$ is the projection of $H_0$ onto the valence space, giving the unperturbed valence energy, and $|\chi_\lambda\rangle$ is the $n$-body wave function which is a superposition of configurations that consist of $n$ nucleons in the valence space. It should be noted that the core nucleus has been considered as the vacuum, and all quantities are relative with respect to the core. The $\hat{Q}$-box can be calculated in terms of power series of $H_1$ perturbatively~\cite{Coraggio2012}, which is the place where approximations are introduced,
\begin{eqnarray}
  \hat{Q}(E)=PH_1P &+& PH_1Q\frac{1}{E-QH_0Q}QH_1P \nonumber\\
                   &+& PH_1Q\frac{1}{E-QH_0Q}H_1\frac{1}{E-QH_0Q}QH_1P \nonumber\\
                   &+& \cdots.
\label{eq:Qbox}
\end{eqnarray}

While the effective Hamiltonian in Eq.~(\ref{eq:secular}) is still energy-dependent, an energy-independent effective Hamiltonian can be obtained by further expanding the denominators in Eq.~(\ref{eq:Qbox}) at the unperturbed energies and regrouping the perturbation diagrams \cite{RevModPhys.49.777,BRANDOW1967}. As a result, all unlinked diagrams are cancelled out in every order, and only linked diagrams remain \cite{RevModPhys.49.777,BRANDOW1967}. In the derivation, a new kind of diagrams called folded diagrams arise. The detailed derivation can be found in Refs. \cite{RevModPhys.49.777,BRANDOW1967}. We call such a perturbation (removing the energy dependence by expanding the energy denominators) the Rayleigh--Schr\"odinger (RS) perturbation. In the RS perturbation, the energy denominators appearing in the diagrams involve only the unperturbed energies.
It is worth noting that the cancellation of unlinked diagrams holds for both degenerate and nondegenerate model spaces~\cite{RevModPhys.49.777,BRANDOW1967} . For a nondegenerate model space, however, the evaluations of the folded diagrams are more complicated \cite{KUO1995205,TAKAYANAGI2014501} than the degenerate case. In the RS perturbation, the energy-independent effective Hamiltonian is
\begin{eqnarray}
  H_{\text{eff}}=PH_0P+\hat{Q}(E^{(\nu)}_0)+\text{folded diagrams},
  \label{eq:Heff_RS}
\end{eqnarray}
where $E^{(\nu)}_0$ is the unperturbed energy. For a nondegenerate model space, $E^{(\nu)}_0$ takes different values for different initial states. For example, when calculating $\langle ab | H_{\text{eff}}| cd \rangle$, $E^{(\nu)}_0$ takes $\epsilon_c+\epsilon_d$. The explicit calculations of the folded diagrams up to a certain order are complicated, especially for a nondegenerate model space \cite{TAKAYANAGI2014501}. Fortunately, the folded diagrams start to appear in the third order expansions. In this paper we calculate the RS perturbation up to second order, not involving any calculations of the folded diagrams.

Anothor way to obtain the energy-independent effective Hamiltonian is to first directly solve the energy-dependent secular equation (\ref{eq:secular}) and then construct the energy-independent effective interaction. We call this method (obtaining the effective interaction by solving the energy-dependent secular equation) the Brillouin--Wigner (BW) perturbation.
The energy-dependent secular equation (\ref{eq:secular}) can be solved self-consistently by iterations. Assuming that we have obtained the eigenvalues and corresponding valence space eigenvectors of Eq.~(\ref{eq:secular}), the energy-independent effective Hamiltonian $H_{\text{eff}}$ defined in the valence space can be constructed as
\begin{eqnarray}
H_{\text{eff}}=\sum\limits_{\lambda=1}^d E^{(\nu)}_\lambda |\chi_\lambda\rangle \langle \widetilde\chi_\lambda|,
\label{eq:Heff}
\end{eqnarray}
where $\langle \widetilde \chi_\lambda|$ is the biorthogonal state of $| \chi_\lambda \rangle$ ( $\langle \widetilde\chi_\lambda| \chi_\mu\rangle =\delta_{\lambda\mu}$), and $d$ is the dimension of the valence configurations. $H_{\text{eff}}$ is usually non-Hermitian or, more accurately, quasi-Hermitian. It can be hermitized by a similarity transformation. The Hermitized Hamiltonian $\bar{H}_{\text{eff}}$ is
\begin{eqnarray}
\bar{H}_{\text{eff}}=\frac{1}{\sqrt{UU^\dagger}} H_{\text{eff}} \sqrt{UU^\dagger},
\end{eqnarray}
where $U$ is the non-unitary matrix in which columns are the valence eigenvectors.

There have already been several iteration methods developed to solve Eq.~(\ref{eq:secular}) and obtain $H_{\text{eff}}$, such as Krenciglowa-Kuo (KK) \cite{KUO1995205,KRENCIGLOWA1974171} iteration and Lee-Suzuki (LS) \cite{doi:10.1143/PTP.64.2091,LEE1980173} method. KK and LS methods are designed for the case with a degenerate model space. There are also extended iteration methods for nondegenerate model spaces \cite{Takayanagi2011a}. Recently, a so-called $\hat{Z}$-vertex was suggested in place of $\hat{Q}$-box when solving Eq. (\ref{eq:secular}) to give a more stable numerical solution of the effective Hamiltonian \cite{PhysRevC.83.024304,DONG20141}. In the original papers using the $\hat Z$-vertex \cite{PhysRevC.83.024304,DONG20141}, the eigen problem, i.e., Eq.~(\ref{eq:secular}) was solved by graphical method.
In this paper, we use the numerical iteration to solve Eqs.~(\ref{eq:HeffE},\ref{eq:secular}) with $\hat{Z}$-vertex replacing $\hat{Q}$-box in the process.
The using of $\hat{Z}$-vertex can avoid possible singularities. In detail, the iteration contains following steps:
\begin{enumerate}
\item give an initial guess of $E^{(\nu)}_\lambda$ and $|\chi_\lambda\rangle$, $\lambda=1,2,...,d$;
\item construct the effective Hamiltonian $H_{\text{eff}}=\sum \limits_{\lambda=1}^d [ PH_0P + \hat{Z}(E^{(\nu)}_\lambda)] |\chi_\lambda\rangle \langle \widetilde\chi_\lambda|$;
\item diagonalize $H_{\text{eff}}$ to obtain new $E^{(\nu)}_\lambda$ and $|\chi_\lambda\rangle$, $\lambda=1,2,...,d$;
\item repeat steps 2 and 3 until a convergence is reached.
\end{enumerate}
Since the $H_{\text{eff}}$ constructed in step 2 is generally non-Hermitian, the diagonalization in step 3 is accomplished by a general matrix diagonalization subroutine. 
In our actual calculations, the energy dependence of the $\hat{Z}$-vertex is quite gentle, and the iteration turns out to be unexpectedly effective.

At the end of the iteration, we obtain the eigen energies $E^{(\nu)}_\lambda$ and the valence-space eigenvectors $|\chi_\lambda\rangle$, as well as the effective Hamiltonian $H_{\text{eff}}$. Let $|\chi_\lambda\rangle$ be normalized to 1, we can further obtain the norm ($N$) of the full wave function and the norm ($N^Q$) of the wave function in the excluded space \cite{Takayanagi2011a},
\begin{eqnarray}
&&N_\lambda=1+N^Q_\lambda,\nonumber\\
&&N^Q_\lambda=-\hat{Q}_1(E^{(\nu)}_\lambda),
\label{eq:rhoq}
\end{eqnarray}
where $\hat{Q}_1$ is the first derivative of $\hat{Q}$-box. The wavefunction probability in the excluded space is $\rho^Q_\lambda=\frac{N^Q_\lambda}{N_\lambda}$. If we have chosen a physically meaningful valence space where the probability $\rho^Q$ is small, the energy dependence of the diagrams would be gentle.

\subsection{Perturbations in the Hartree--Fock Basis}
The intrinsic Hamiltonian of an $A$-nucleon system can be written as
\begin{eqnarray}
  H&=&\sum\limits_i^A{\frac{{\bm p}_i^2}{2m}} + \sum\limits_{i<j=1}^A V_{ij} - \frac{(\sum\limits_i^A{{\bm p}_i})^2}{2mA}\nonumber\\
  &=&(1-\frac{1}{A})\sum\limits_i^A{\frac{{\bm p}_i^2}{2m}} + \sum\limits_{i<j=1}^A\left(V_{ij}-\frac{{\bm p}_i\cdot {\bm p}_j}{mA}\right),
  \label{eq:H}
\end{eqnarray}
where ${\bm p}_i$ is the nucleon momentum in laboratory and $V_{ij}$ is the two-body $NN$ interaction.
 
We first performed a spherical HF calculation in the HO basis \cite{Hu2016}, obtaining the HF s.p. energies and wave functions. In the HF basis, the Hamiltonian~(\ref{eq:H}) can be written as
\begin{eqnarray}
  H&=&H_0+H_1\nonumber\\
   &=&\sum\limits_i \epsilon_i\hat{a}_i^\dagger \hat{a}_i + \left [ \sum\limits_{i<j=1}^A \left(V_{ij}-\frac{{\bm p}_i\cdot {\bm p}_j}{mA} \right) - \sum\limits_i^A U_i^{\text{HF}} \right ],
  \label{eq:Hhf}
\end{eqnarray}
where $\epsilon_i$ and $\hat{a}_i^\dagger$ are the energy and the creation operator of the HF s.p. state labelled by $i$, respectively. $U_i^{\text{HF}}$ is the self-consistent HF potential obtained in the iteration. $H_0=\sum\limits_i \epsilon_i\hat{a}_i^\dagger \hat{a}_i$ is the unperturbed Hamiltonian and $H_1$ is the perturbation term.

Valence nucleons occupy the HF s.p. orbits in the valence space, e.g., $0d_{5/2,3/2}$ and $1s_{1/2}$ for the {\it sd} shell which we are interested in the present paper.
The two-body matrix elements in the  HF basis are obtained by basis transformation, and they will be used in the $\hat{Q}$-box calculations.
One can find the detailed perturbation diagrams in Refs. \cite{Coraggio2012,Hjorth-Jensen1995}. Because the self-consistent HF auxiliary potential is taken, the ($V-U^{\text{HF}}$)-insertion diagrams are cancelled out \cite{Coraggio2012}. Note that the HF iteration is performed for the core nucleus, i.e., $^{16}$O for the {\it sd} shell. 

In the RS perturbation, the energy-independent effective Hamiltonians for systems with one and two valence particles can be directly obtained by Eq. (\ref{eq:Heff_RS}). We calculate the RS perturbation up to second order so that no folded diagrams are involved. While in the BW perturbation, to derive the effective Hamiltonians, we need to solve Eq.~(\ref{eq:secular}) for systems with one and two valence particles.
For one-valence-particle system with $A_{\text c}+1$ nucleons, Eq.~(\ref{eq:secular}) becomes
\begin{eqnarray}
\sum\limits_j \left [ \epsilon_i\delta_{ij} + \hat{S}_{ij}(E^{(\nu)}_\lambda) \right ] \chi_j^\lambda&=&E^{(\nu)}_\lambda \chi_i^\lambda,
\label{eq:secular1B}
\end{eqnarray}
where $\hat{S}$ denotes the one-body $\hat{Q}$-box diagrams. For two-valence-particle system with $A_{\text c}+2$ nucleons,
Eq.~(\ref{eq:secular}) becomes
\begin{eqnarray}
\sum\limits_{kl}\left [ (\epsilon_i +\epsilon_j)\delta_{ij,kl} + \hat{Q}_{ijkl}(E^{(\nu)}_\lambda) \right] \chi_{kl}^\lambda&=&E^{(\nu)}_\lambda \chi_{ij}^\lambda,
\label{eq:secular2B}
\end{eqnarray}
where the indexes $i, j, k, l$ label the HF s.p. states in the valence space. In the present work, the $\hat{Q}$-box is calculated up to third order. The $(A_\text{c}+1)$- and $(A_\text{c}+2)$-body problems are reduced to the one- and two-body problems, respectively. $\chi_i^\lambda$ (or $\chi_{ij}^\lambda$) is the valence-space superposition coefficients of the $\lambda$-th one-body (or two-body) wave function, and is actually proportional to the experimentally measurable one-body (or two-body) spectroscopic amplitude \cite{Kuo1981},
\begin{eqnarray}
  \chi_i^\lambda (1+N^{Q}_\lambda)^{-1/2} &=& \langle \Psi_0^{A_\text{c}} | \hat{a}_i | \Psi_\lambda^{A_\text{c}+1} \rangle,
\end{eqnarray}
or
\begin{eqnarray}
\chi_{ij}^\lambda (1 + N^{Q}_\lambda)^{-1/2}& =& \langle \Psi_0^{A_\text{c}} | \hat{a}_j\hat{a}_i | \Psi_\lambda^{A_\text{c}+2} \rangle,
\end{eqnarray}
where $| \Psi_0^{A_\text{c}} \rangle$ is the ground state of the core, and $| \Psi_\lambda^{A_\text{c}+1} \rangle$ and $| \Psi_\lambda^{A_\text{c}+2} \rangle$ are the states for the $A_\text{c}+1$ and $A_\text{c}+2$ systems, respectively. $(1+N^{Q}_\lambda)^{-1/2}$ is the normalization factor introduced in Eq.~(\ref{eq:rhoq}). Eqs. (\ref{eq:secular1B}) and (\ref{eq:secular2B}) are solved by employing the numerical iteration mentioned above.

Both in the RS and the BW calculations, the one-body and two-body Hermitized effective Hamiltonians $\bar{H}_{\text{eff}}^{(1)}, \bar{H}_{\text{eff}}^{(2)}$ can be obtained. $\bar{H}_{\text{eff}}^{(1)}$ gives the valence s.p. energies, which are quantities that can be directly compared with the experimental data. The two-body effective interaction $V_{\text{eff}}$ can be obtained as $V_{\text{eff}}=\bar{H}_{\text{eff}}^{(2)}-\bar{H}_{\text{eff}}^{(1)}$. The effective s.p. energies (SPEs) and the effective interaction $V_{\text{eff}}$ can be further used as the input of the SM calculations for other {\it sd}-shell nuclei.


In the language of RS perturbation, the iteration in BW perturbation is equivalent to summing up a series of folded-diagrams. In our BW perturbation calculations, we calculate the $\hat{Q}$-box up to third order, while the iteration equivalently sums up some RS-perturbation folded-diagrams up to infinite order. It's not clear whether the partial summation of the diagrams can offer a better approximation. The RS perturbation is size-extensive (i.e., scale properly with the size of the system) in each order \cite{MBbook}, while the partial summation in BW perturbation breaks this property, which may be a problem for the cases with many valence nucleons. On the other hand, contributions from some high-order unlinked folded diagrams are included by the iteration. These contributions should have be cancelled out by the non-folded unlinked diagrams of the same order \cite{RevModPhys.49.777} which are absent in our calculation where $\hat{Q}$-box is calculated up to third order.

It's interesting to make an analogy between perturbation methods with the EOM-CC method \cite{PhysRevC.88.024305,PhysRevLett.113.142502}.
Up to third order perturbation, the intermediate 2-particle (2p), 3-particle-1-hole (3p1h), 4-particle-2-hole (4p2h) states out of the model space are shown in the two-body diagrams. Therefore, the correlations of the Q-space 2p, 3p1h, 4p2h configurations are taken into account perturbatively for the system with two valence nucleons. Actually, for example, the wave function for a state of $^{18}$O in our $sd$-shell calculation can be formally written as
\begin{eqnarray}
  |\Psi_\lambda^{^{18}\text{O}} \rangle &=& \left(1+ \frac{1}{E_\lambda^{(\nu)}-QHQ}QH_1P\right)\sum\limits_{ij} \chi_{ij}^\lambda \hat{a}_i^\dagger\hat{a}_j^\dagger |^{16}\text{O} \rangle
\end{eqnarray}
where $i, j$ run over valence neutron states. $\chi_{ij}^\lambda$ is the component in the valence space obtained from Eq.~\ref{eq:secular2B} and $|^{16}\text{O} \rangle$ is the formal ground state of $^{16}$O. The second term in the bracket shows as the intermediate states in the diagram calculations in Eq.~(\ref{eq:Qbox}), thus gives the Q-space 2p, 3p1h, 4p2h components when the diagrams are truncated up to third order. In Refs. \cite{PhysRevC.83.054306,PhysRevC.88.024305} where EOM-CC method is introduced, the ground state of the closed-shell nucleus is first solved within the coupled-cluster theory, based on which the states of the open-shell nuclei are described in the same way as MBPT by considering 2p and 3p1h excitations. The difference is that in EOM-CC, the states of the open-shell nuclei are obtained by diagonalization, thus the correlations are taken into account nonperturbatively. In the MBPT formalism, the core nucleus is separated out and there is no need to first solve the core.

In the application to the $sd$ shell, we can enlarge the valence space to $sdpf$ to see the influences to the spectra. The effect of the $pf$ shell, which is originally taken into account perturbatively by particle-particle ladder-diagrams, now is taken into account nonperturbatively by SM diagonalization.

\section{Results and Discussions}
We investigate the {\it sd} shell using the chiral N$^3$LO potential \cite{Entem2003,Machleidt2011a}. The potential is softened using the $V_{\text{low-}k}$ technique \cite{Bogner2003} with a cutoff $\Lambda=2.5$ fm$^{-1}$. We take the frequency parameter $\hbar\omega=22$ MeV for the HO basis and truncate the basis with $N_{\text{max}}=2n+l=10$. Within the HO basis, we perform the spherical HF calculations and perturbation calculations. We compare the results  with valence-space IM-SRG calculations \cite{HERGERT2016165,Tsukiyama2012} with the same input Hamiltonian. The IM-SRG calculations are performed using the code from \cite{IM-SRG-code}. The valence space that consists of particle states $0d_{5/2,3/2}$ and $1s_{1/2}$ has been chosen unless otherwise stated.
\subsection{Convergence Analyses}
\begin{table}
  \caption{\label{table:sp} Calculated $sd$-shell effective s.p. energies with a cutoff $\Lambda=2.5$ fm$^{-1}$ $V_{\text{low-}k}$ N$^3$LO potential. BW-2nd and BW-3rd indicate the HF BW perturbation calculations in which $\hat Q$-box is calculated up to second and third order, respectively. RS-2nd indicates the HF RS perturbation up to second order. For comparison, we also present the IM-SRG calculations and experimental data. We take the underlying HO basis parameter $\hbar\omega=22$ MeV and basis truncation $N_{\text
{max}}=10$. Energies are in MeV.}
  \begin{ruledtabular}
    \begin{tabular}{ccccccc}
                &    HF   &  RS-2nd    &  BW-2nd  &  BW-3rd    &  IM-SRG       & Expt.\\\hline
$\nu 0d_{5/2}$  &   1.89  &    -4.73   &  -4.36   &  -4.55     &   -4.89       &  -4.14    \\
$\nu 1s_{1/2}$  &   1.75  &    -3.30   &  -3.03   &  -2.95     &   -3.23       &  -3.27    \\
$\nu 0d_{3/2}$  &   7.38  &    1.83    &  2.23    &  2.53      &   2.53        &  0.94     \\
$\pi 0d_{5/2}$  &   5.45  &    -0.80   &  -0.47   &  -0.57     &   -0.93       &  -0.60    \\
$\pi 1s_{1/2}$  &   4.96  &     0.31   &  0.54    &  0.67      &   0.40        &  -0.10    \\ 
$\pi 0d_{3/2}$  &   10.48 &     5.39   &  5.72    &  6.06      &   6.04        &  4.40     \\
    \end{tabular}
  \end{ruledtabular}
\end{table}

\begin{figure}[!htbp]
\centering
\includegraphics[width=0.8\columnwidth]{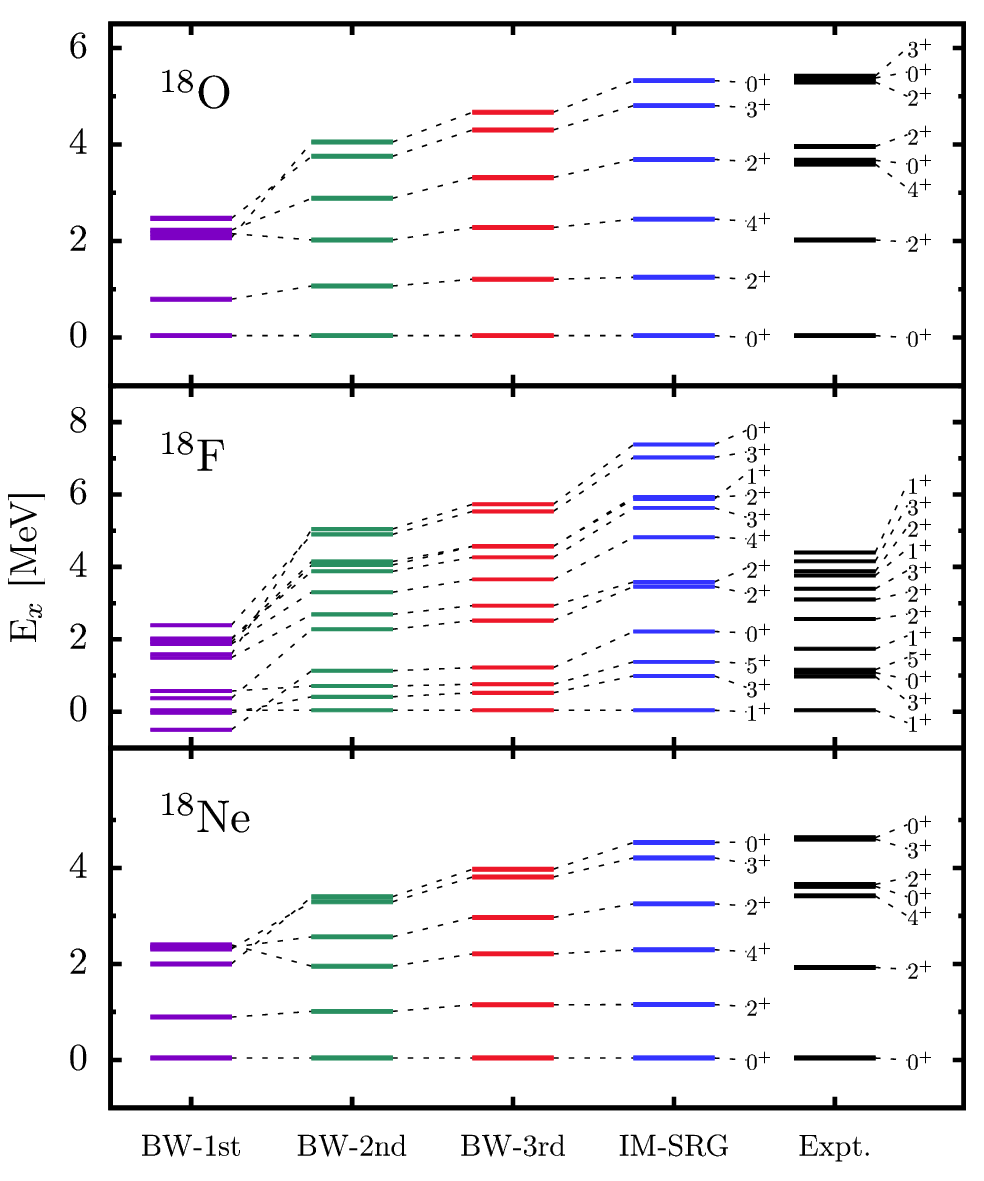}
\caption{\label{fig:converge} Calculated spectra of $^{18}$O, $^{18}$F and $^{18}$Ne by the HF BW perturbations in which $\hat Q$-box is calculated up to first, second and third order, compared with the nonperturbative IM-SRG calculations with the same $\Lambda=2.5$ fm$^{-1}$ $V_{\text{low-}k}$ N$^3$LO potential. The valence space ${0d_{5/2,3/2}, 1s_{1/2}}$ has been chosen. The underlying HO basis parameter $\hbar\omega=22$ MeV is taken and the basis is truncated with $N_{\text{max}}=10$. The experimental data are taken from \cite{TILLEY19951}.}
\end{figure}

In Table \ref{table:sp}, we show the s.p. energies obtained from the calculations. The effective s.p. energies can correspond to experimental single-particle states excited in $^{17}$O and $^{17}$F with respect to the ground state of $^{16}$O. From Table \ref{table:sp}, we see the good convergence. The BW perturbation with $\hat Q$-box up to second order has already given well converged s.p. energies. Even without the three-body nuclear force in our calculations, the obtained s.p. energies and the experimental data are in reasonable agreement. Figure~\ref{fig:converge} displays the calculated spectra of the two-valence-particle nuclei $^{18}$O, $^{18}$F and $^{18}$Ne with the BW perturbation where $\hat Q$-box is calculated up to first, second and third order. The main purpose of the present work is to benchmark the HF-basis perturbation calculations with the nonperturbative IM-SRG calculations. The BW-3rd calculated s.p. energies are close to the IM-SRG calculations within a discrepancy of 0.3 MeV. The discrepancies between the BW-3rd and IM-SRG calculations in spectrum shown in Fig.~\ref{fig:converge} for $^{18}$O and $^{18}$Ne are within 0.6 MeV. In $^{18}$F the discrepancy can reach as large as 1.6 MeV. From the perspective of RS perturbation, a reason for the discrepancy may be that the iteration in BW perturbation equivalently sums up some higher-order unlinked folded diagrams. The unlinked diagrams should have been cancelled out by the non-folded unlinked diagrams of the same order which are absent in our calculations where $\hat{Q}$-box is calculated up to third order. We will discuss the RS perturbation calculations in next subsection.

\begin{figure}[!htbp]
\centering
\includegraphics[width=0.8\columnwidth]{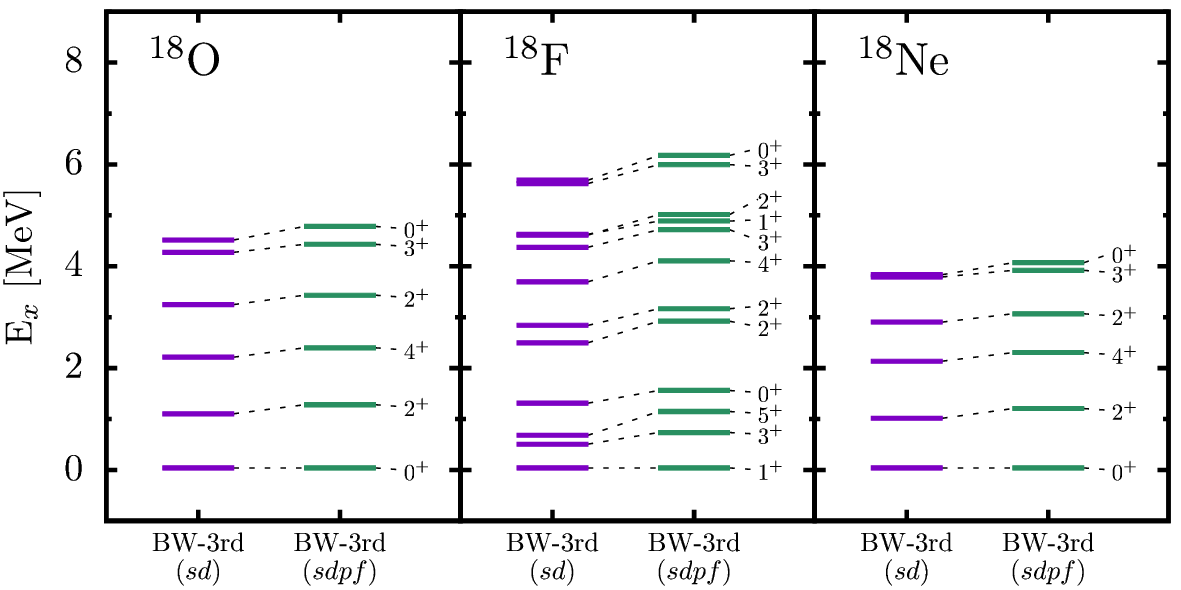}
\caption{\label{fig:BW_sd_sdpf} The HF-basis BW-3rd spectrum calculations of $^{18}$O, $^{18}$F and $^{18}$Ne nuclei with the model space $sd$ and the extended model space $sdpf$. $\Lambda=2.5$ fm$^{-1}$  $V_{\text{low-}k}$ N$^{3}$LO potential is used. The underlying HO basis parameter $\hbar\omega=22$ MeV is taken and the basis is truncated with $N_{\text{max}}=10$.}
\end{figure}

In order to discuss the convergence against the model space, we have performed the BW-3rd calculations with a larger model space $sd+pf$, shown in Fig.~\ref{fig:BW_sd_sdpf}. The $pf$-shell effect that was taken into account by particle-particle ladder perturbation diagrams in the calculations with the $sd$ model space, now is included explicitly in the diagonalization. We see that the calculations with the valence space $sd$ and $sdpf$ are nearly the same. This agreement provides a test of the accuracy of our perturbation calculations, showing that the calculation which perturbatively takes into account the correlations out of the model space should be a good approximation. In fact, the norm out of the model space $\rho^Q$ introduced in Eq.~(\ref{eq:rhoq}) are at most 10\% in our calculations for $^{18}$O, $^{18}$F, $^{18}$Ne. The low-lying states of these nuclei are indeed dominated by the configurations in the $sd$ space.

\subsection{Comparison between BW and  RS Perturbations}
\begin{figure}[!htbp]
\centering
\includegraphics[width=0.8\columnwidth]{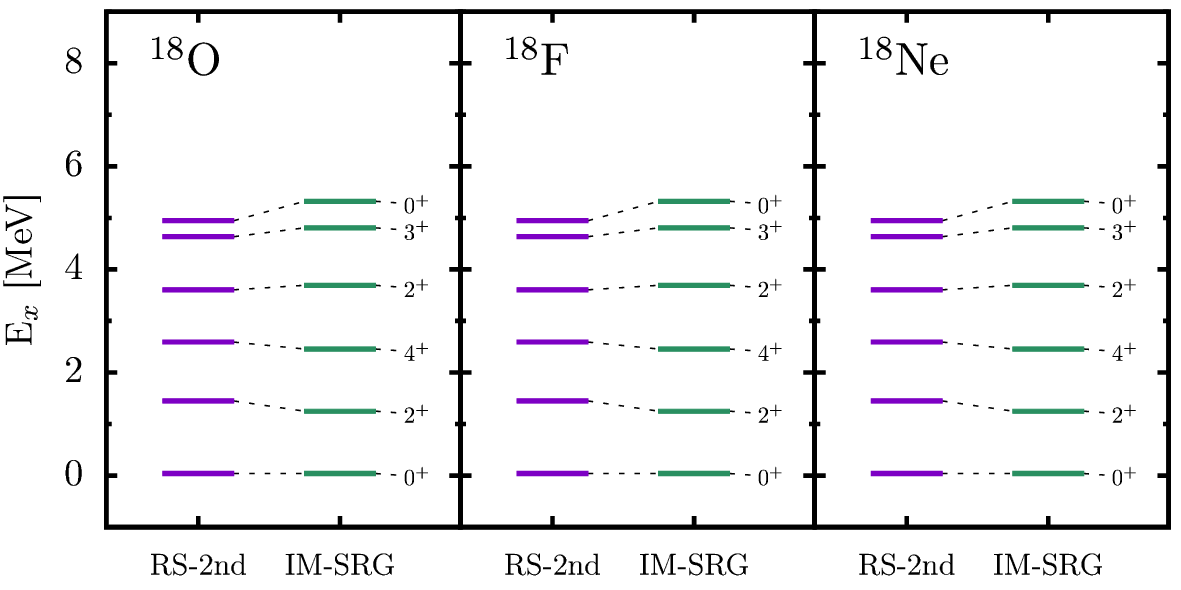}
\caption{\label{fig:RS_imsrg}Calculated spectra of $^{18}$O, $^{18}$F and $^{18}$Ne by the HF RS perturbation up to second order, compared with the nonperturbative IM-SRG calculations with the same $\Lambda=2.5$ fm$^{-1}$ $V_{\text{low-}k}$ N$^3$LO potential. The valence space ${0d_{5/2,3/2}, 1s_{1/2}}$ has been chosen. The underlying HO basis parameter $\hbar\omega=22$ MeV is taken and the basis is truncated with $N_{\text{max}}=10$.}
\end{figure}

\begin{figure}[!htbp]
\centering
\includegraphics[width=0.8\columnwidth]{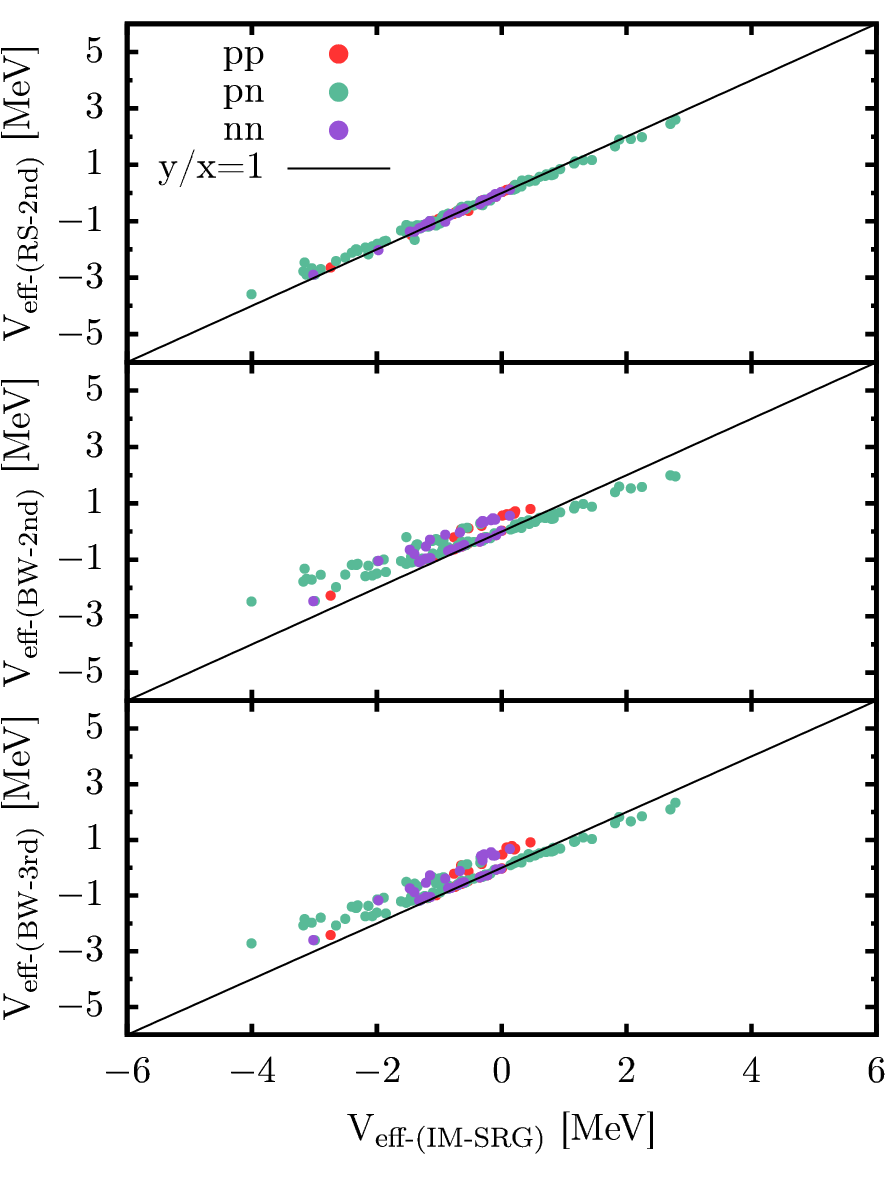}
\caption{\label{fig:compareV} $sd$-shell proton-proton (pp), proton-neutron (pn) and neutron-neutron (nn) effective interaction obtained from the HF BW second-order, third-order and RS second-order calculations, compared with that obtained from IM-SRG. In all the calculations, $V_{\text{low-}k}$ at cutoff $\Lambda=2.5$ fm$^{-1}$ starting from the chiral two-body N$^{3}$LO potential is used. The underlying HO basis parameter $\hbar\omega=22$ MeV is taken and the basis is truncated with $N_{\text{max}}=10$.}
\end{figure}

In Fig.~\ref{fig:RS_imsrg}, we show the spectra obtained from the RS second-order calculations, compared with that obtained from IM-SRG. Compared with BW-3rd shown in Fig.~\ref{fig:converge}, RS-2nd is in better agreement with IM-SRG. Actually, the discrepancy of RS-2nd and IM-SRG is within 0.5 MeV for all the states shown in Fig.~\ref{fig:RS_imsrg}. Given the other uncertainties of the nuclear {\it ab initio} calculations, this small discrepancy is quite satisfactory. The second-order RS perturbation can be performed with low computational cost, thus can be applied to heavy nuclei. The calculated effective s.p. energies and effective two-body interaction can be further used as the input of the SM to calculate other $sd$-shell nuclei. We compare the obtained effective two-body interaction matrix elements in Fig.~\ref{fig:compareV}. An overall better agreement with IM-SRG is found for RS-2nd. The agreement of the matrix elements between RS-2nd and IM-SRG is encouraging. With this effective interaction, the calculations of other $sd$-shell nuclei would be close to the results of IM-SRG.

\section{Summary}
We apply the HF-basis perturbation calculations of realistic effective interactions to the $sd$ shell. Two types of perturbations are performed: BW perturbation and RS perturbation. It is shown that the HF basis provides good convergences for the perturbation calculations. We also benchmark the calculations with the nonperturbative method IM-SRG with the same input Hamiltonian. The perturbation calculations are in fairly good agreement with IM-SRG. From the comparison, we find that the HF RS perturbation gives even better results compared with the BW perturbation. In fact, the discrepancy between the second-order RS perturbation and IM-SRG calculations in low-lying spectrum for $^{18}$O, $^{18}$F and $^{18}$Ne is within 0.5 MeV. The HF realistic effective interactions derived can be further used as the input of the SM to calculate other $sd$-shell nuclei. We conclude that, with simple formalism and low computational cost, the perturbation framework based on the HF basis using soft nuclear forces can be an efficient and reliable tool for the first principle studies of open-shell nuclei.

\begin{acknowledgments}
  This work has been supported by the National Natural Science
  Foundation of China under Grants No. 11235001, No. 11320101004 and No.
  11575007; and the CUSTIPEN (China-U.S. Theory Institute for Physics with
  Exotic Nuclei) funded by the U.S. Department of Energy, Office of Science
  under Grant No. DE-SC0009971.
\end{acknowledgments}

\bibliography{effVHF}

\end{document}